\documentclass[epj]{svjour}
\usepackage{amsmath}
\usepackage{amssymb}
\usepackage{graphicx}
\usepackage[usenames]{color}
\usepackage{soul}
\usepackage{ulem}
\usepackage{multirow}

\newcommand{\JPB}{J. Phys. B}
\newcommand{\NJP}{New J. Phys.}

\newcommand{\PL}{Phys. Lett.}
\newcommand{\RMP}{Rev. Mod. Phys.}
\newcommand{\cpc}{Comp. Phys. Commun.}

\begin{document} 

\title{Effect of optical lattice potentials on the vortices in rotating dipolar Bose-Einstein condensates}
\titlerunning{Effect of optical lattice potentials on the vortices in dipolar BECs}
\author{R. Kishor Kumar \and P. Muruganandam
}                     
\authorrunning{R. Kishor Kumar and P. Muruganandam}
\offprints{}          
\institute{School of Physics, Bharathidasan University, Palkalaiperur Campus, Tiruchirappalli 620024, Tamilnadu, India}
\date{Received: date / Revised version: date}
%
\abstract{We study the interplay of dipole-dipole interaction and optical lattice (OL) potential of varying depths on the formation and dynamics of vortices in rotating dipolar Bose-Einstein condensates. By numerically solving the time-dependent quasi-two dimensional Gross-Pitaevskii equation, we analyse the consequence of dipole-dipole interaction on vortex nucleation, vortex structure, critical rotation frequency and number of vortices for a range of OL depths. Rapid creation of vortices has been observed due to supplementary symmetry breaking provided by the OL in addition to the dipolar interaction. Also the critical rotation frequency decreases with an increase in the depth of the OL. Further, at lower rotation frequencies the number of vortices increases on increasing the depth of OL while it decreases at higher rotation frequencies. This variation in the number of vortices has been confirmed by calculating the rms radius, which shrinks in deep optical lattice at higher rotation frequencies.
} 
\maketitle

\section{Introduction}

The experimental realization of dipolar Bose-Einstein condensates (BECs) of bosonic atoms interacting via long range and anisotropic dipole-dipole interaction has created a new insight in the understanding of the physics of cold dipolar atoms and molecules~\cite{Griesmaier2005}. The recent progress in the study of dipolar BECs has exposed various fascinating physics due to the peculiar competition between an isotropic, short-range contact interaction and an anisotropic, long-range dipolar interaction. The significant features of dipolar BECs are the emergence of biconcave shaped ground state structures, stability dependence on trap geometry, roton-like dip in the dispersion relation, and structured cloud featuring a $d$-wave symmetry during collapse~\cite{Dutta2007,Koch2008,Saito2009}. 

Very recently much attention has been given to understand the properties of dipolar BEC in optical lattice and in multilayer systems~\cite{Trefzger2011,Baranov2012}. Optical lattice (OL) is a spatially periodic potential realized in experiments using standing waves of counter propagating laser beams~\cite{Bloch2005}. Dipolar BEC in OL has proven to be a more suitable candidate for simulating condensed matter systems with long range anisotropic interactions in a controllable environment. There are studies on dipolar BECs in OL from within as well as beyond mean-field description~\cite{Trefzger2011}. Dipolar BECs in OL are of high relevance in condensed matter physics due to the appearance of insulating metastable states, Mott-insulator phase, checkerboard supersolid phase, strongly correlated regime, localization in disordered lattice and in random potential~\cite{Goral2002}. 

An interesting property of Bose-Einstein condensates is the creation of quantized vortices due to excitation. In experiments, excitation of a BEC has been achieved either by rotating magnetic traps or by laser stirring. Vortices are usually formed above a critical rotation frequency. There are several studies on vortices in dipolar BECs using mean-field models~\cite{Yi2006,Baranov2008,Klawunn2008,Abad2009,Malet2011,Kishor2012,Van2007}. A second-order like phase transition of straight and helical vortex lines occurs due to the influence of dipolar orientation has been reported~\cite{Klawunn2008}. It has been shown that the dipolar BEC strongly influence the number, structure and stability of vortices. Further, in dipolar BECs the critical rotation frequency for vortex nucleation found to decrease as the strength of dipolar interaction increases~\cite{Malet2011,Kishor2012}. It has also been realized that the dipolar interaction increases the number of vortices while the contact interaction enhances the vortex stability~\cite{Kishor2012}.

Earlier studies on BEC vortices in OL have been mainly focussed on conventional Bose gas with local and isotropic interaction. In BEC experiments, the condensate is loaded into a static OL and then studied by applying rotation~\cite{Tung2006,William2010}. Conventional BECs in OL under rotation have been shown to exhibit various interesting properties such as vortex structures, vortex structural phase transition, pinning effect of vortices with the peak of shallow OL, vortex lattice in the deep OL~\cite{Tung2006,William2010,Reijnders2004,Pu2005,Yasunaga2007,sato2007,Daniel2008,Kato2011}. However, no attempt has been made so far to explore the influence of dipolar interaction on BEC vortices in OL. 

In the present paper we focus on the study of vortices in rotating dipolar BEC in OL. In particular, we investigate the influence of dipolar strength and the depth of the OL on the formation and dynamics of vortices. The presence of OL has been found to stimulate rapid vortex formation in dipolar BEC. Also the critical rotation frequency decreases in the presence of OL. By calculating the expectation value of angular momentum we notice that the presence of OL suppresses the shape deformation during the development of vortices. Further, ordered vortices have been created in dipolar BEC due to the pinning of vortices within the peaks of OL. Depending upon the strength of dipolar interaction, phase transition from regular pentagon structure to square and triangular vortex lattice has been observed. We calculate the number of vortices as a function of rotation frequency for different OL depths and contact interaction strengths in dipolar BECs and compare them with that of conventional (non-dipolar) BECs. In addition, we find that there is a reduction in the number of vortices at higher rotation frequencies for stronger dipolar BECs in OL. We also calculate the rms radius of the dipolar BEC as a function of OL depth as well as rotation frequencies for different contact interaction strengths. 

The present paper is organized as follows. In Sec.~\ref{sec:frame} we provide an overview on the mean field Gross-Pitaevskii equation describing the properties of a rotating dipolar BEC confined in an axially symmetric harmonic trap potential and OL. In Sec.~\ref{sec:dyn}, we present the numerical studies on the formation and dynamics of vortices in dipolar BECs of $^{52}$Cr, $^{168}$Er and $^{164}$Dy atoms. We analyze the formation of vortices in a pure dipolar BECs due to the presence of OL. We also calculate the critical rotation frequency and number of vortices for a range of OL depths. Then, in Sec.~\ref{depth}, we study the characteristic features of these vortices in the presence of contact interaction by calculating the rms radii as a function of rotation frequency and OL depth. Finally, in Sec.~\ref{sec:conclusion} we provide a summary and conclusion. 

\section{Theoretical description}
\label{sec:frame}
The dynamics of rotating BECs can be studied using mean field Gross-Pitaevskii (GP) equation~\cite{Fetter2009,Adhikari2002,Bao2006}. At absolute zero temperature a dipolar BEC with $N$ atoms, each of mass $m$, loaded in OL in a rotating frame can be described by the Gross-Pitaevskii equation as~\cite{Yi2006}
\begin{align}  
i \frac{\partial \phi({\bf r},t)}{\partial t}  = &  \left[ -\frac{1}{2}\nabla^2 +V({\bf r}) 
+ 4\pi a N\vert \phi({\bf r},t)\vert^2 - \Omega L_z \right. \nonumber \\ & \left. 
+  N \int U_{dd}({\bf r -r'})\vert\phi({\bf r'},t)\vert^2d^3{ r'}
\right] \phi({\bf r},t), \label{eqn:dgpe} 
\end{align} 
where $V({\bf r})=V_{ho}({\bf r})+V_{OL}(\rho)$, $\rho \equiv (x,y)$ is the confining axially symmetric harmonic potential and optical lattice potential, $\phi({\bf r},t)$  the wave function at time $t$ with normalization $\int \vert\phi({\bf r},t)\vert^2 d {\bf r}=1$, and $a$ is the atomic scattering length. The axial and radial trap frequencies of the harmonic potential, $V_{ho}({\bf r})$, are $\omega_z$ and $\omega_\rho$, respectively, and are related to the trap aspect ratio as $\lambda=\omega_z/\omega_\rho$. In equation~(\ref{eqn:dgpe}) length is measured in units of harmonic oscillator length $l \equiv \sqrt{\hbar/m\omega_\rho}$, frequency in units of $\omega_\rho$, time $t$ in units of $\omega_\rho^{-1}$. $L_z= - i (x\partial_y-y \partial_x)$ corresponds to the $z$-component of the angular momentum due to the rotation of the dipolar BEC about $z$ axis with angular velocity $\Omega$. Here $\Omega$ is expressed in units of the radial trap frequency $\omega_\rho$. The integral term in equation~(\ref{eqn:dgpe}) accounts for the dipole-dipole interaction with
\begin{align}
U_{dd}({\mathbf  x})=a_{dd}\frac{1-3\cos^2 \theta}{\vert {\mathbf x}\vert ^3},
\end{align}
where ${\mathbf x}= {\mathbf r} -{\mathbf r'}$ determines the relative position of dipoles and $\theta$ is the angle between ${\mathbf x}$ and the direction of polarization, $z$. 
The constant $a_{dd}=\mu_0\bar{\mu}^2 m/(12\pi\hbar^2)$  is a length characterizing the strength of dipolar interaction and, its experimental value for $^{52}$Cr, $^{168}$Er and $^{164}$Dy are $16a_0$, $66a_0$ and $131 a_0$, respectively, where $a_0$ is the Bohr radius~\cite{Youn2010}. $\bar{\mu}$ corresponds to the magnetic dipole moment of a single atom and $\mu_0$ the permeability of free space. 

The dimensionless three-dimensional harmonic trap and two-dimensional optical lattice is given by
\begin{align}
 V({\bf r}) =  \frac{1}{2} \rho^2 + \frac{1}{2} \lambda^2 z^2 + V_0 \left[\sin^2(kx)+\sin^2(ky)\right],
\end{align} 
where ${\bf r}\equiv (\vec\rho,z)$, with $\vec\rho$ the radial coordinate and $z$ the axial coordinate, $V_0$ is the depth of the OL and $k$ is the wave number.

In pancake-shaped traps the side by side arrangement of dipoles provides the necessary repulsive dipole-dipole interaction to stabilize the dipolar BECs. On the other hand, polarized dipoles align in a head-to-tail configuration in cigar-shaped traps and provide an attractive dipolar interaction, which leads to collapse. The use of strong pancake trap helps the experimental realization of dipolar BEC with zero scattering length~\cite{Koch2008}. Excitations near the instability regime in weak pancake trap leads to angular collapse of dipolar BEC~\cite{Wilson2009}. We consider a highly oblate dipolar BEC with trap aspect ratio $\lambda=100$ for our present study. 
{In this  case the dipolar BEC is assumed to be in the ground state, 
\begin{eqnarray}
\phi_{1D}(z) = \frac{1}{(\pi d_z^2)^{1/4}}\exp\left(-\frac{z^2}{2d_z^2}\right), 
\;\; \omega_z d_z^2=1, \label{eqn:zfun}
\end{eqnarray}
of the axial trap so that the wave function $\phi(\mathbf r)$ can be written as, 
\begin{eqnarray}
\phi(\mathbf r) = \phi_{1D}(z)  \phi_{2D}(x,y),\label{eqn:ans}
\end{eqnarray}
where $\phi_{2D}(x,y)$ is the 2D wave function and $d_z = 1/\sqrt{\lambda}$.
The dynamics of the rotating dipolar BEC can effectively be studied in two-dimensions 
\begin{figure*}[!ht]
\begin{center}
\includegraphics[width=0.80\linewidth,clip]{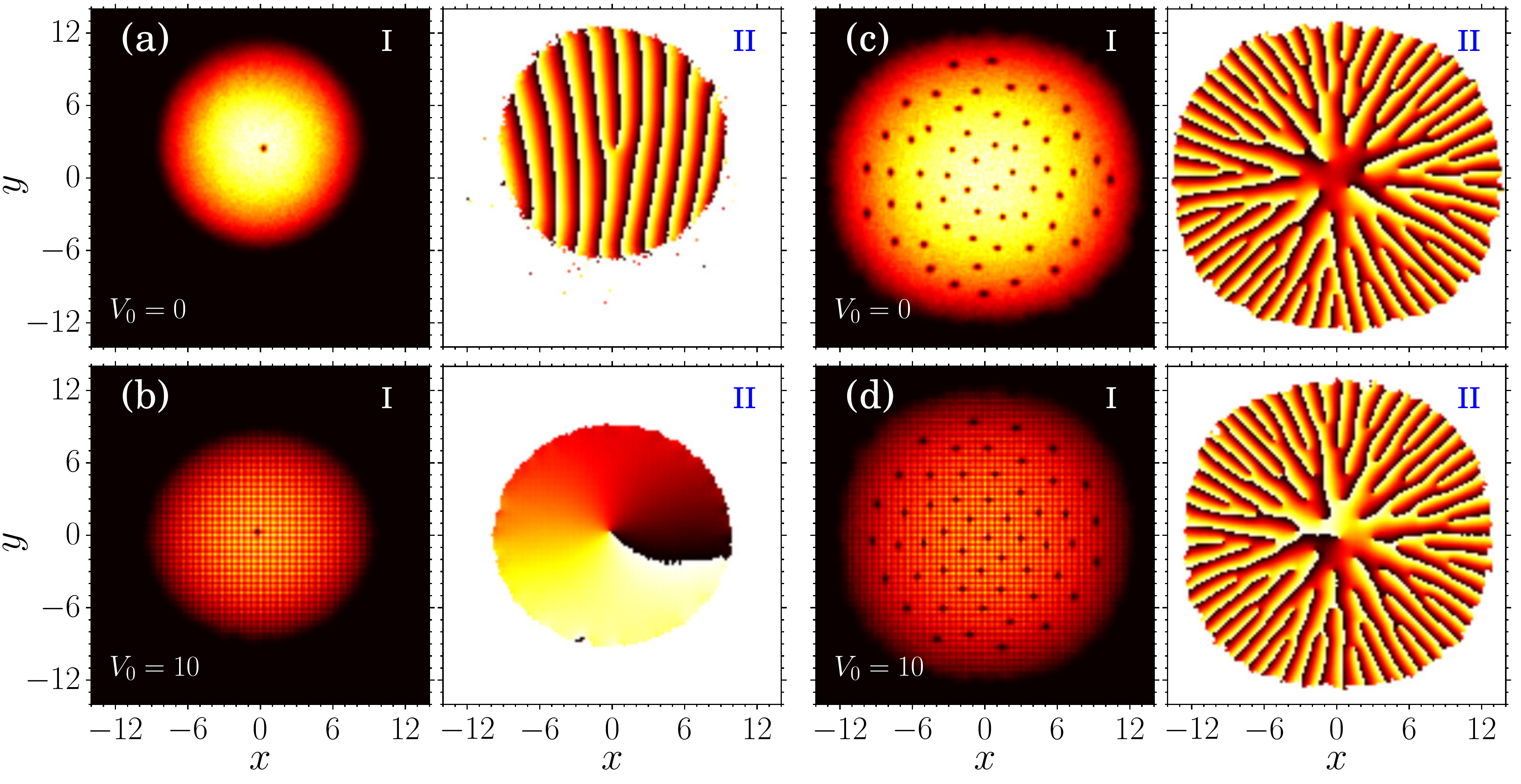}
\end{center}
\caption{(Color online) Contour plots of densities, $\vert \phi_{2D} \vert^2$, showing the comparison of  vortices in the absence and in the presence of OL for a strongly dipolar BEC of $^{164}$Dy atoms with $a=0$ and $a_{dd} = 131\,a_0$: (a) single off-centered vortex at the critical rotation frequency, $\Omega = 0.275$, (b) single well centered vortex at $\Omega = 0.258$ in the presence of OL, (c) slightly distorted vortex lattice structure in the absence of OL, and (d) ordered vortex lattice in the presence of OL for $\Omega = 0.7$. The right panels (II) show the phase patterns of the wave function $\phi_{2D}$.}
\label{fig1}
\end{figure*}
by simply integrating out the $z$ dependence in equation~(\ref{eqn:dgpe}), as~\cite{Pedri2005,Fisch2006,Muruganandam2012},}
\begin{align}
i\frac{\partial \phi_{2D}(\vec \rho,t)}{\partial t}  = & \left[-\frac{\nabla_\rho^2}{2}
+ V_{2D} -\Omega L_z +\frac{4\pi aN}{\sqrt{2\pi}d_z}\vert \phi_{2D}(\vec \rho,t)\vert ^2 
 \right.\nonumber \\ & 
+ \frac{4\pi a_{dd}N}{\sqrt{2\pi}d_z} \int \frac{d^2k_{\rho}}{(2\pi)^2} \mbox{e}^{i\bf {k_\rho} \cdot \vec \rho} \, 
{\tilde n}({\bf k_\rho})  \notag \\ & \left. \times 
h_{2D}\left(\frac{k_\rho d_z}{\sqrt{2}} \right) \right]  
\phi_{2D}\left(\vec \rho,t \right). \label{red2d}
\end{align}
where $V_{2D} = \left(x^2+y^2\right)/2 + V_0 \left[\sin^2(kx)+\sin^2(ky)\right]$ is the two-dimensional harmonic trap and optical lattice potential. In equation~(\ref{red2d}),
$\tilde n({\bf k_\rho}) = \int \exp(i{\bf k_\rho. \vec\rho)} \vert\phi_{2D}(\vec\rho)\vert^2 d\vec\rho$, $k_\rho\equiv (k_x,k_y)$, $h_{2D}(\xi)=2-3\sqrt{\pi}\xi \exp (\xi^2)\, \mbox{erfc}(\xi)$, and the dipolar term is written in Fourier space. The lattice spacing ($d_{lat}$) and amplitude ($V_0$) of optical lattice potential can be varied by tuning the frequency and intensity of laser. We have chosen the OL spacing as $d_{lat}= \lambda_L/2 \approx534$nm, where $\lambda_L=1064nm$ is the wavelength of the laser used in experiments~\cite{Muller2011}. The corresponding dimensionless parameters for OL spacing, $\tilde d_{lat}$ = $ \pi/k = 0.534$ and $k=1.87 \pi$.

It may be noted that in a highly oblate trap and with perpendicular polarization of dipoles the dipolar interaction potential ($1/\vert r\vert^3$) can be effectively treated as a short range and isotropic. The short range potentials can be defined by $s$-wave scattering length $a$ and the condensate properties (such as energy, chemical potential) can be explained interms of the gas parameter $n a^D$, where $D=2$ and $n = \vert \phi_{2D} \vert^2$ is the condensate density. However, due to the larger spatial extension of the dipolar interaction the universal description becomes precise for the larger values of gas parameter for dipolar potential than for the usual short range potentials~\cite{Astrakharchik2008}. Hence one cannot treat the dipolar interaction in 2D to an  equivalent effective repulsive contact interaction as the condensates properties in both the cases are different.

In the following, we study the formation of vortices in $^{52}$Cr, $^{168}$Er and $^{164}$Dy condensates in OL by solving the two dimensional GP equation~(\ref{red2d}). For this purpose, we numerically solve equation~(\ref{red2d}) using a combined split-step Crank-Nicolson and fast Fourier transform (FFT) based numerical scheme~\cite{Muruganandam2012,Muruganandam2009}.  For the present study we fix the number of atoms as $N=10\,000$. All the numerical simulations in this  manuscript are carried out with $dx=dy=0.2$ (space step) and $dt=0.004$ (time step). 

\section{Formation of vortices in pure dipolar BEC in optical lattice}
\label{sec:dyn}

We prepare the ground state wavefunction by solving equation~(\ref{red2d}) numerically using imaginary time propagation in the presence of both harmonic and optical lattice potentials but without rotation ($\Omega=0$). This ground state is then allowed to evolve with real time propagation by including rotation ($\Omega \ne 0$). A phenomenological dissipation is included to facilitate the smooth vortex formation~\cite{Tsubota2002,Kasamatsu2005,Garcia-Ripoll2001}. The dissipation is introduced by replacing `$i$' with `$(i-\gamma)$' in the time dependent equation (\ref{red2d}), where $\gamma$ $(\sim 10^{-5})$ accounts for the strength of dissipation. The solution of the GP equation with the dissipative term is reliable with the collective damped oscillations of the condensate. Moreover, the vortex lattice correspond to local minimum of the total energy in the configuration space and the transition from a non-vortex state to a vortex state requires energy dissipation. When the condensate begins to rotate the surface becomes unstable and ripples are developed on the surface as time progress. Then these ripples gradually improve into vortices and reach a stable configuration at a finite time. Further these vortices are pinned within the peaks of optical lattice potential. First we compare the formation of vortices in a dipolar BEC of $^{164}$Dy atoms ($a_{dd} = 131 a_0$ and $a=0$) both in the absence and in the presence of OL.
\begin{figure*}[!ht]
\begin{center}
\includegraphics[width=0.80\linewidth,clip]{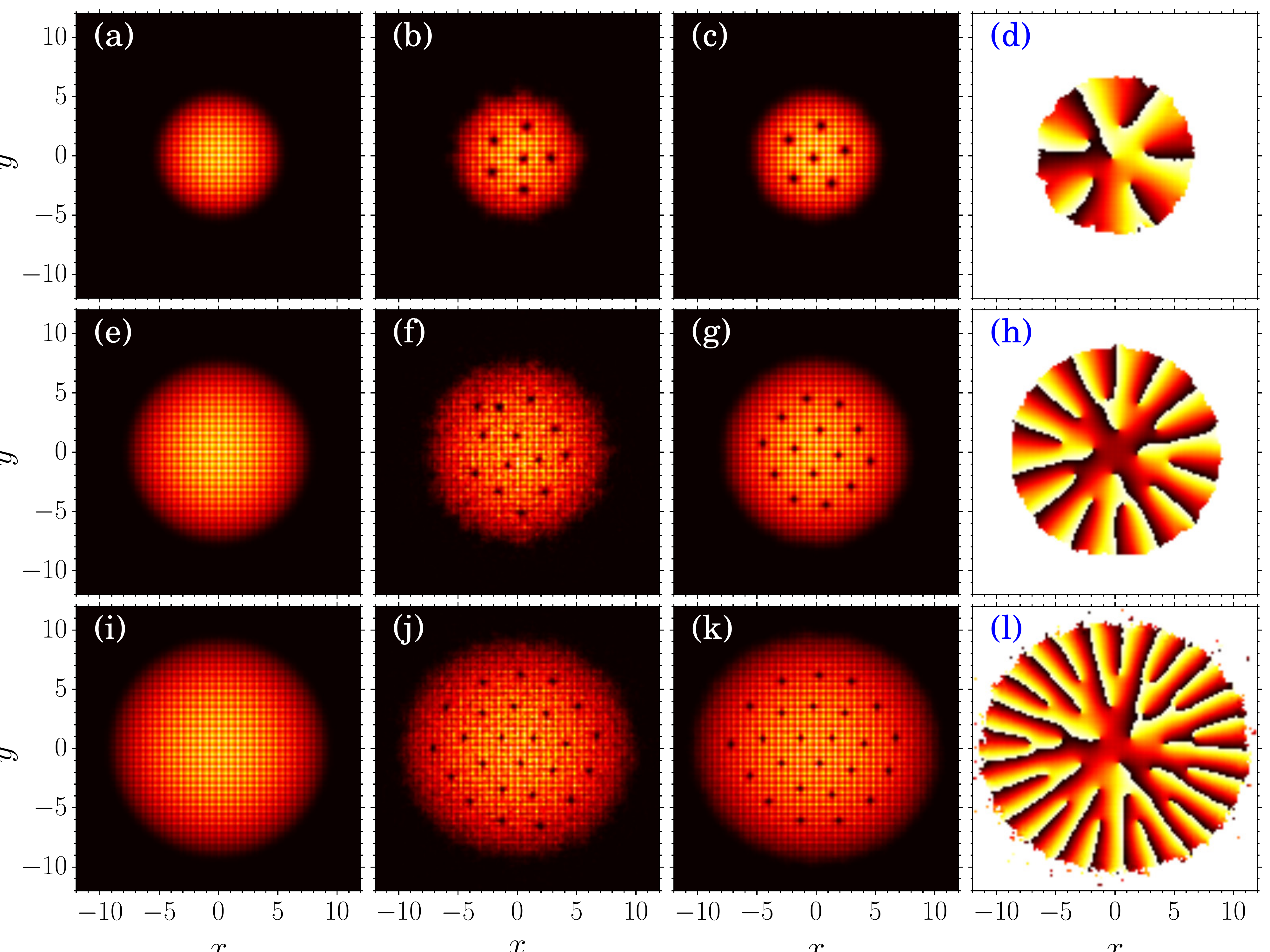}
\end{center}
\caption{(Color online) Contour plots of the densities, $\vert \phi_{2D}(\rho, t) \vert^2$, showing the development of vortices in dipolar BEC in OL for  $a_{dd} = 16\, a_0$: (a) $t=0$ (b) $t=24\, 000\, \omega_\rho^{-1}$,  (c) $t=36\, 000\, \omega_\rho^{-1}$, for  $a_{dd} = 66\, a_0$ (e) $t=0$, (f) $t=2400\, \omega_\rho^{-1}$ (g) $t = 8000\, \omega_\rho^{-1}$, and for $a_{dd} = 131\, a_0$ (i) $t=0$, (j) $t=120\, \omega_\rho^{-1}$ (k) $t = 3000\, \omega_\rho^{-1}$.  The phase distributions of the equilibrium state are shown in (d), (h) and (l). The other parameters are fixed at $V_0  = 10$, $a = 0$, $N = 10\,000$, $\lambda = 100$ and $\Omega  = 0.5$.}
\label{fig2}
\end{figure*}
In the absence of OL, a single off-centered vortex as shown in Figure~\ref{fig1}(a) is created at the critical rotation frequency, $\Omega = \Omega_\text{crit} = 0.275$. While a well centered single vortex as in Figure~\ref{fig1}(b) is formed when the OL with depth $V_0=10$ is introduced. The right panels (II) show the phase patterns of corresponding wave functions of Figures~\ref{fig1}(a) and ~\ref{fig1}(b). The phase varies continuously from $0$ (dark) to $2\pi$ (bright) and the location of the vortices are clearly visible at the branching (bifurcation) point. At a higher rotation frequency, say for example $\Omega = 0.7$, an equilibrium state with $39$ vortices as in Figure~\ref{fig1}(c) is formed in the absence of OL. On the other hand, a well ordered pattern of $38$ vortices in square lattice [see Figure~\ref{fig1}(d)] is observed when applying OL. 

Next we study the dynamics during the formation of vortices in BECs of $^{52}$Cr, $^{168}$Er and $^{164}$Dy atoms. In Figure~\ref{fig2} we show the snapshots of the density contours $\vert \phi_{2D} \vert^2$ and phase patterns for OL depth $V_0=10$ and rotation frequency $\Omega=0.5$. Here we compare the time taken for the formation of steady state vortices for different dipolar strengths. Figures~\ref{fig2}(a)-(c) show the development of vortices for $a_{dd} = 16\, a_0$ ($^{52}$Cr BEC). In this case, the surface ripples are formed at $t \sim 12\,000\, \omega_\rho^{-1}$ and an equilibrium state of  $6$ vortices pinned within the peaks of OL at $t \sim 36\, 000\, \omega_\rho^{-1}$ [Figure~\ref{fig2}(c)]. The corresponding phase pattern of final wavefunction is shown in Figure~\ref{fig2}(d). 

Figures~\ref{fig2}(e)-(g) show the development of vortices for $a_{dd} = 66\, a_0$ ($^{168}$Er BEC). Here the surface ripples are formed at time $t \sim 120\,\omega_\rho^{-1}$, which is much faster when compared to that of $^{52}$Cr BEC discussed above. Further, an equilibrium state of $14$ vortices as shown in Figure~\ref{fig2}(g) is formed. The phase profile of the final wave function is shown in Fig~\ref{fig2}(h). 

We have also shown the snapshots of vortices for $^{164}$Dy BEC ($a_{dd} = 131\, a_0$) in Figures~\ref{fig2}(i)-(k). Here the surface ripples are formed more rapidly, that is at $t \sim 24\,\omega_\rho^{-1}$, and a very stable pattern with $24$ vortices are created [Figure~\ref{fig2}(k)] for $t > 2000\,\omega_\rho^{-1}$. Figure~\ref{fig2}(l) depicts the phase profile of the final wave function. 

The above observation clearly indicates that the time taken for the creation of steady state vortices in dipolar BECs in OL decreases considerably with the increase of dipolar strength. To understand the influence of OL, we have also estimated the approximate time ($t_{vor}$) for the formation of steady state of vortices in a dipolar BEC in the absence of OL ($V_0 = 0$) with the same rotation frequency ($\Omega = 0.5$). 
\begin{table}[!ht]
\caption{The approximate time ($t_{vor}$) for the creation of equilibrium vortices and the number of vortices ($N_v$) in the presence and absence of OL for pure dipolar BEC ($a = 0$) with $\Omega = 0.5$.}
\label{table1}
\begin{center}
\begin{tabular}{c|c|c|c|c}
\hline
\multirow{2}{*}{$\displaystyle\frac{a_{dd}}{a_0}$} & \multicolumn{2}{c|}{$t_{vor}$ ($\omega_\rho^{-1}$)} &\multicolumn{2}{c}{$N_v$}\\
\cline{2-5}
      & $V_0=0$ & $V_0=10$ & $V_0=0$&  $V_0=10$ \\
\hline
$16$  &$44\,000$ & $16\,000$ &  $5 $   &  $6$  \\
$66$  &$41\,000$ & $4\,000$  &  $12$   & $14$  \\
$131$ &$39\,000$ & $2\,000$  &  $23$   & $24$  \\
\hline  
\end{tabular}
\end{center}
\end{table} 
These times are presented in Table~\ref{table1} along with the number of vortices ($N_v$) for different dipole-dipole interaction strengths ($a_{dd}$) in the presence as well as absence of OL. Actually in the absence of OL it takes a very long time $(\sim 40\, 000 \, \omega_\rho^{-1})$ while it reduces drastically, about few thousands $\omega_\rho^{-1}$ (one order less), with the presence of OL for strongly dipolar BECs. It is easy to see from the above results that the creation of steady state vortices in rotating dipolar BECs in OL is much rapid when the dipolar strength is large. 

The faster nucleation of vortices is common in BECs of strongly dipolar atoms. This is because of the fact that the dipole-dipole interaction breaks the axial symmetry more easily and speed up the formation of vortices~\cite{Malet2011}. Faster creation of vortices in the absence of OL has been observed in rotating dipolar BEC (about one order less time) when compared to conventional condensate~\cite{Kishor2012}.  It may be noted that in $^{164}$Dy BEC the vortices manifest quickly and the condensate has a larger rms radius, when compared to $^{52}$Cr and $^{168}$Er BECs, due to the strong dipolar interaction. The optical lattice provides a supplementary symmetry breaking in addition to the dipole-dipole interaction, which actually stimulates the rapid creation of steady state vortices.  One may note that in conventional BECs with pure s-wave (contact) interaction the critical rotation frequency for the vortex nucleation is independent of interaction strength even though the vortex nucleation depends on the existence of two-body interactions. While in the case of dipolar BEC the critical frequency is strongly affected by the magnitude of dipole-dipole interactions. 

It is also worth to discuss the orderliness of the vortex patterns in the dipolar BEC in OL. In conventional BECs phase transitions of vortex lattice from Abrikosov vortex lattice to the pinned lattice and rich variety of vortex structures have been reported~\cite{Pu2005,sato2007}. The phase transitions in vortex structures are studied as functions of strength, density of the OL and the interaction among the vortices~\cite{Pu2005,sato2007}. In this connection, different patterns in the equilibrium vortex structures of dipolar BEC in OL have been observed. For example, regular pentagon structure with one vortex at center, square and triangular vortex lattice structures with slight distortion as shown in Figures~\ref{fig2}(c), \ref{fig2}(g) and \ref{fig2}(k), for $^{52}$Cr, $^{168}$Er and $^{164}$Dy atoms, respectively, are evident.

\begin{figure}[!ht]
\begin{center}
\includegraphics[width=0.80\columnwidth,clip]{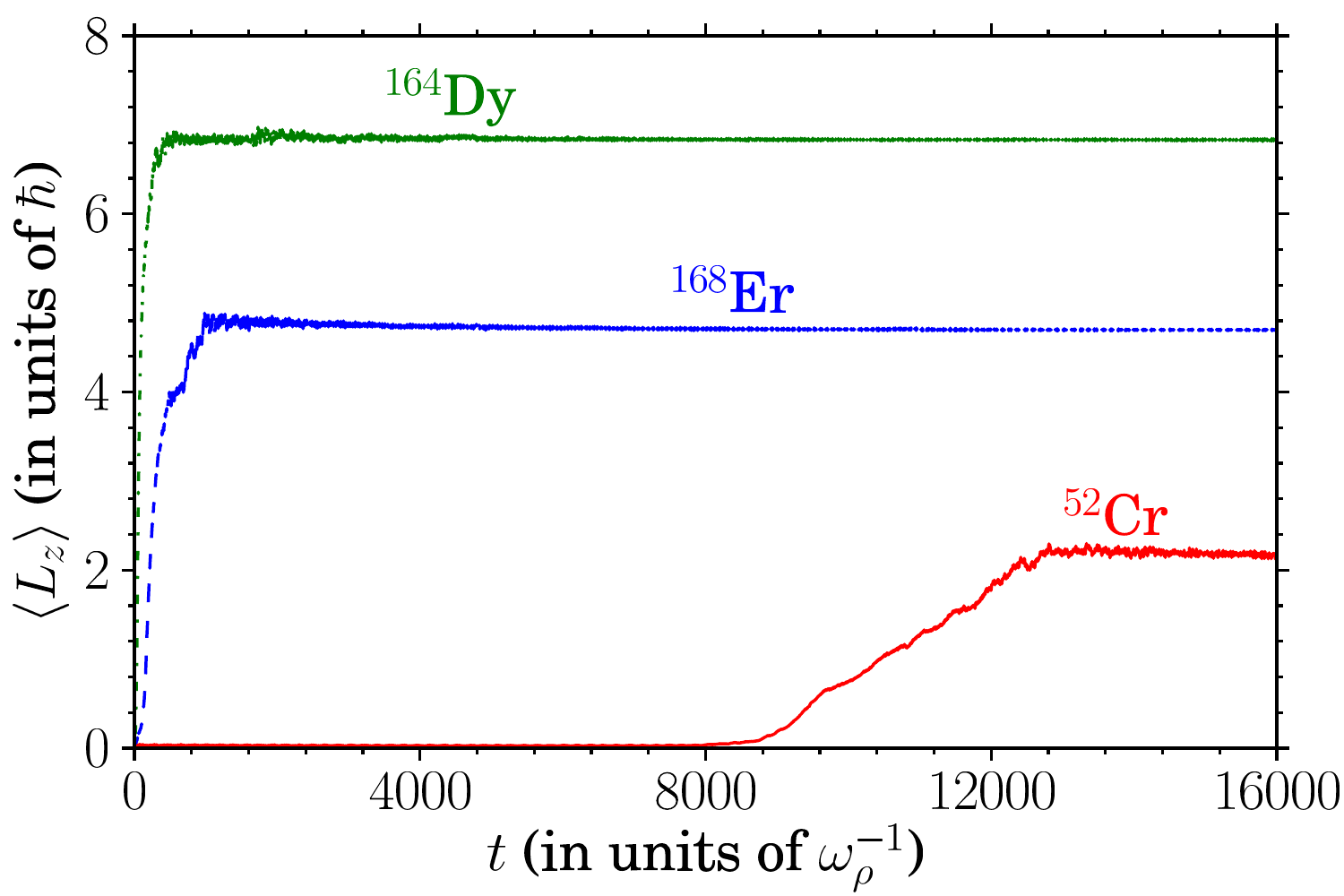}
\end{center}
\caption{(Color online) Time evolution of the expectation value of angular momentum, $\langle L_z \rangle$, during the development of vortices shown in Figure~\ref{fig2}.}
\label{fig3}
\end{figure}

We also note that the shape deformation and quad\-rupole oscillations are suppressed due to the presence of OL. A rotating dipolar BEC in a harmonic trap normally shows shape deformation and quadrupole oscillations during the development of vortices~\cite{Kishor2012}. This can be easily seen by studying time evolution of angular momentum, which normally shows the quadrupole oscillations with large amplitudes. Here we study the time evolution of angular momentum for the cases discussed above in Figure~\ref{fig2} by calculating the expectation value of angular momentum defined as
\begin{align}
\langle L_z \rangle = i \int \phi^{\star}(\vec \rho, t) (y\partial_x-x\partial_y) \phi(\vec \rho, t)\, d\vec\rho.
\end{align}
In Figure~\ref{fig3} we plot $\langle L_z \rangle $ as a function of time for $^{52}$Cr, $^{168}$Er and $^{164}$Dy atoms. 
$\langle L_z \rangle$ gradually increases and finally settles to a steady value after a long time ($t \approx 13\,000\, \omega_\rho^{-1}$) for $^{52}$Cr BEC. On the other hand the expectation value of angular momentum rapidly increases soon after applying rotation in $^{168}$Er and $^{164}$Dy BECs then settles quickly ($t \approx 1\, 000\, \omega_\rho^{-1}$ and $t \approx 120\, \omega_\rho^{-1}$, respectively) to a steady value confirming the stable vortex pattern.

\section{Effect of varying optical lattice depth and influence of contact interaction on the vortices in dipolar BEC}
\label{depth}

Next, it is of interest to study the effect of varying OL depth on the vortices in dipolar BEC. In conventional BECs the presence of OL has been found to reduce the critical rotation frequency for vortices and creates more number of vortices with respect to the depth of the OL~\cite{Kato2011}.  It may be noted that a pure dipolar BEC, when the contact interaction is made zero, easily collapses in deep OL. However, stable rotating dipolar BEC can be formed in moderate depths of OL.

\begin{figure*}[!ht]
\begin{center}
\includegraphics[width=0.80\linewidth,clip]{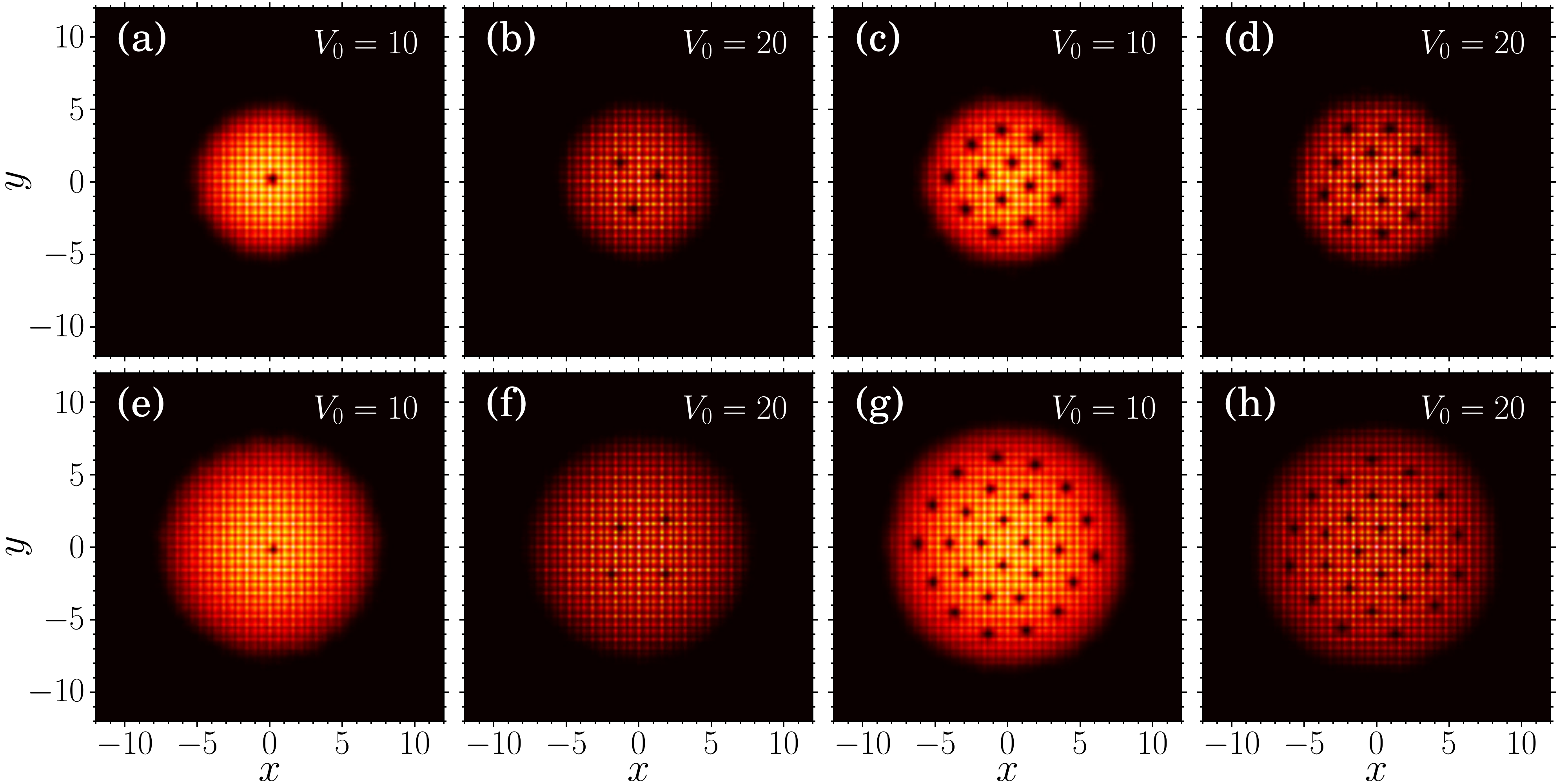}
\end{center}
\caption{(Color online) Contour plots of the density $\vert \phi_{2D} \vert^2$ showing steady state vortices in a rotating dipolar BEC with  $a=0$, $N = 10\,000$, $\lambda = 100$. For  $a_{dd}=16 \,a_0$, (a) and (b) $\Omega  = 0.385$ , (c) and (d) $\Omega = 0.7$. For $a_{dd}=66 \,a_0$  (e) and (f)  $\Omega = 0.295$, and (g) and (h) $\Omega = 0.7$}
\label{fig4}
\end{figure*}

We calculate the number of vortices $N_v$ for different OL depths, for example, $V_0 =  10$ and $20$. For a pure dipolar BEC of $^{52}$Cr ($a=0$ and $a_{dd} = 16 \, a_0$) in OL with depth $V_0=10$ a single vortex is formed as shown in Figure~\ref{fig4}(a) at a critical rotation frequency, $\Omega_c  \approx 0.385$. 
When the depth of the OL is increased to $V_0=20$ three vortices are found as shown in Figure~\ref{fig4}(b) for the same rotation frequency. While at a higher rotation frequency, for example at $\Omega=0.7$, the number of vortices remains the same ($N_v = 13$) for different OL depths. In Figure~\ref{fig4}(c) and \ref{fig4}(d) we show the density profiles of rotating $^{52}$Cr BEC with OL depths $V_0=10$ and $20$, respectively. Similar behavior has been observed in the case of $^{168}$Er BEC at lower rotation frequencies. For example, at $\Omega_c=0.295$, a single stable vortex as shown in Figure~\ref{fig4}(e) appears for OL depth $V_0=10$, while four vortices as in Figure~\ref{fig4}(f) are generated for $V_0=20$. However, at a higher rotation frequency $\Omega=0.7$ about $28$ vortices [Figure~\ref{fig4}(g)] appear for $V_0=10$ and $N_v=27$ (one less) for $V_0=20$ as in Figure~\ref{fig4}(h). 

\subsection{Rotation frequency versus number of vortices}

Now we study the dependence of number of vortices $N_v$ on the rotation frequency $\Omega$. The equilibrium number of vortices for a given rotation frequency is proportional to the radius of the rotating superfluid and can be estimated using Feynman's rule as,
\begin{align}
N_v=\frac{m\Omega}{\hbar} \, R_\rho^2(\Omega),
\end{align}
where $m$ is the mass, $\hbar$ the reduced Planck's constant and $R_\rho(\Omega)$ is the radius. Condensates of larger radius can accommodate more number of vortices~\cite{Kishor2012,Fetter2009}. In the absence of OL, the radius $R_\rho(\Omega)$ can be approximated in terms of $R_\rho(0)$. 
\begin{figure*}[!ht]
\begin{center}
\includegraphics[width=0.80\linewidth,clip]{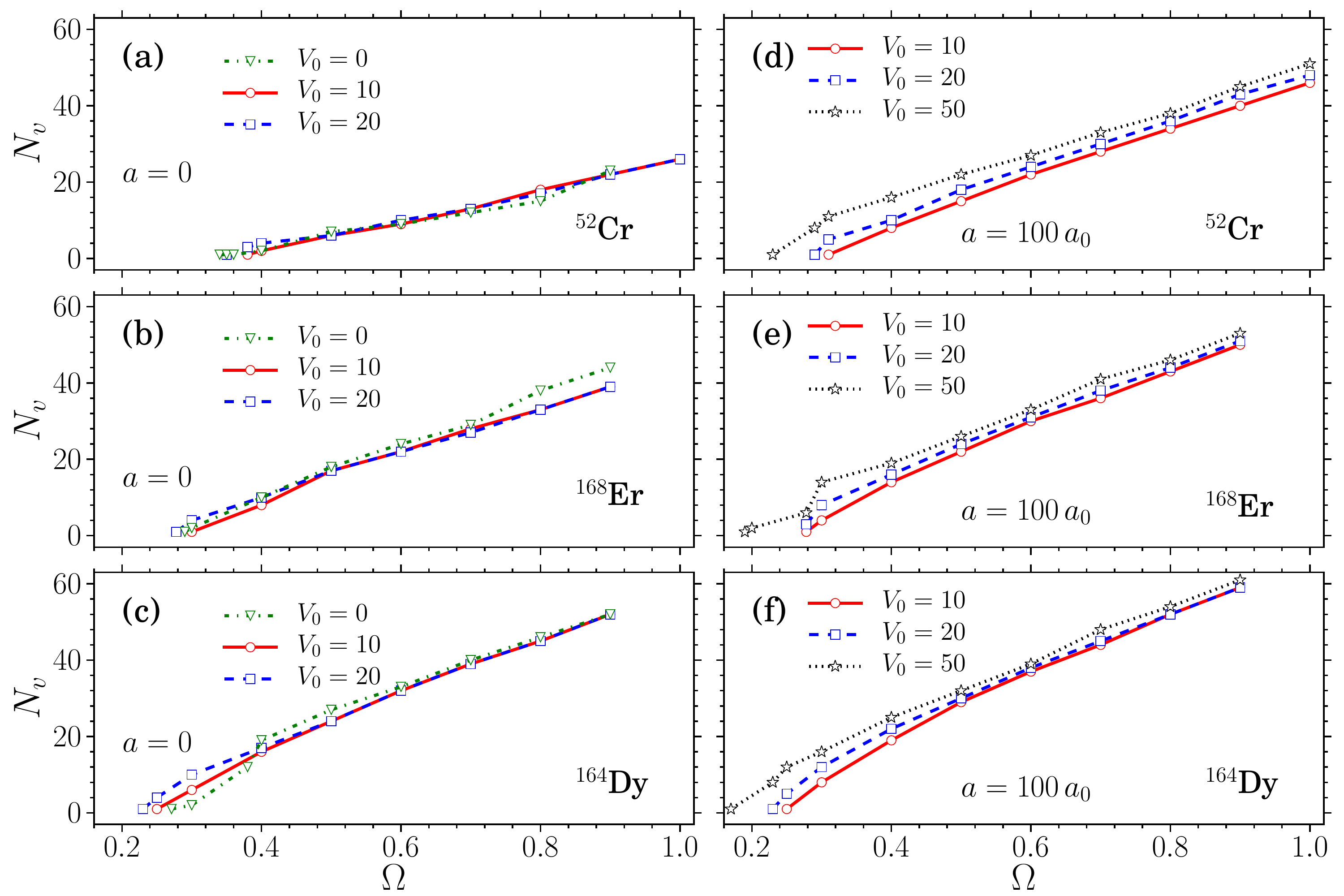}
\end{center}
\caption{(Color online) Plot of the numerically computed equilibrium   number of vortices ($N_v$) as a function of the rotational frequency $\Omega$ for a pure dipolar BECs ($a = 0$) with (a) $a_{dd} = 16\, a_0$, (b) $a_{dd} = 66\, a_0$  and (c) $a_{dd} = 131\, a_0$. Figures (d), (e) and (f) depict the variation of $N_v$ as a function of $\Omega$ with repulsive contact interaction, $a = 100 \, a_0$, for different dipolar strengths as given in (a), (b) and (c), respectively.} 
\label{fig5}
\end{figure*}
However, in the presence of OL this relation is not known. So we numerically analyze the dependence of number of vortices $N_v$ on the rotation frequency.  We have calculated the number of vortices for dipolar BECs as a function of OL depth ($V_0$) and rotation frequency ($\Omega$). 
\begin{figure}[!ht]
\begin{center}
\includegraphics[width=\linewidth,clip]{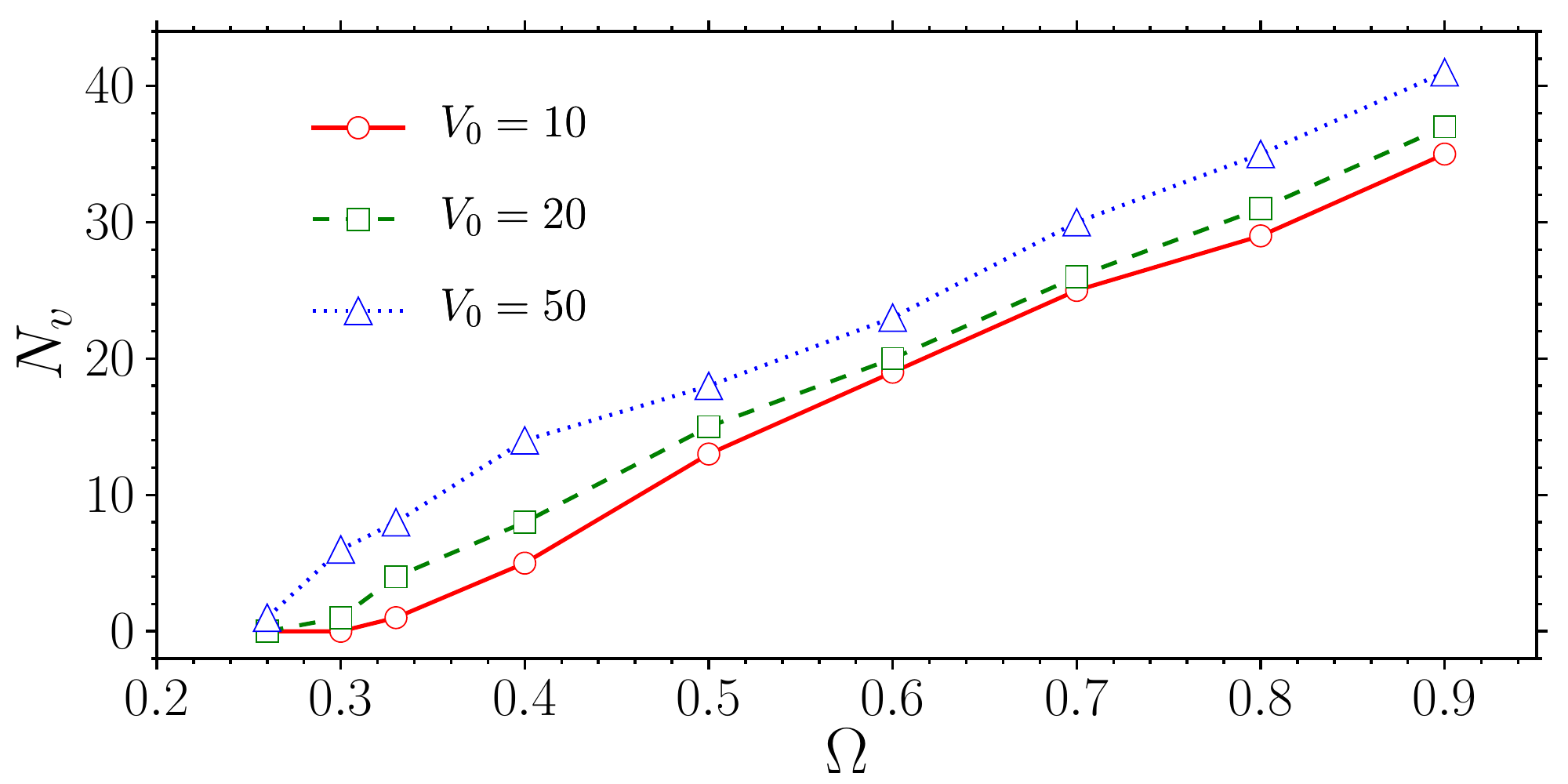}
\end{center}
\caption{(Color online) Plot of the numerically computed equilibrium number of vortices ($N_v$) as a function of the rotational frequency $\Omega$ for conventional (non-dipolar) BEC with $a = 100\, a_0$.} 
\label{fig5a}
\end{figure}
In Figures~\ref{fig5}(a)-\ref{fig5}(c), we plot the equilibrium number of vortices ($N_v$) against the rotation frequency ($\Omega$) for pure dipolar BECs of $^{52}$Cr, $^{168}$Er and $^{164}$Dy atoms, respectively, with different OL depths. For rotation frequencies up to $\Omega \approx 0.5$, the number of vortices varies appreciably with OL depths. However, $N_v$ decreases slightly with OL depths for $\Omega > 0.5$. We have also calculated $N_v$ in the absence of OL and are shown by green/dotted lines with down triangle symbols in Figures~\ref{fig5}(a)-\ref{fig5}(c). Figures~\ref{fig5}(b)-\ref{fig5}(c) for the cases of $^{168}$Er and $^{164}$Dy BECs. These numbers are relatively small at lower rotation frequencies and are large at higher $\Omega$. This suggests that, in a pure dipolar BEC, increasing the depth of the OL favours more number of vortices at lower rotation frequencies ($\Omega<0.5$). However, at higher rotation frequencies the number of vortices decreases slightly with the increase of the OL depth. {To compare the effect of dipolar interaction, the number of vortices as a function of OL depth is calculated for conventional (non-dipolar) BECs and is shown in Figure~\ref{fig5a}. Here $N_v$ increases as the depth of OL for all rotation frequencies.} 

It may be noted that, in a shallow OL potential, atoms can easily tunnel to neighbouring lattice sites and condensate atoms in the lattice site (onsite) do not control the behaviour of the condensate~\cite{Trefzger2011}. On the other hand, condensate atoms in deep OL attain the regime of Josephson-junction array, where fractions of the condensate are well localized at minima of the OL potential~\cite{Kato2011}. As a consequence the overlap of wave function between neighbouring lattice sites (intersite) becomes insignificant and the onsite interaction becomes dominant. In a pure dipolar BEC in deep OL the onsite interaction becomes attractive which destabilizes the condensate and lead to collapse. This collapse may be prevented by the inclusion of suitable repulsive contact interaction.

We have calculated $N_v$ with the inclusion of repulsive contact interaction ($a \neq 0$) in dipolar BECs in the presence of OL. The number of vortices found to increase for all rotation frequencies as the depth of the OL is increased. In Figures~\ref{fig5}(d)-\ref{fig5}(f), we show the variation of $N_v$ as a function of $\Omega$ for different OL depths for BECs of $^{52}$Cr, $^{168}$Er and $^{164}$Dy atoms with $(a = 100 a_0)$. The rotating dipolar BEC is found to be stable for $V_0 > 40$ due to the presence of contact interaction, while pure dipolar BEC becomes unstable when $V_0 > 40$. In all the cases of dipolar BEC with $a \neq 0$ considered here, the equilibrium  $N_v$ increases with the increase of OL depth for a wide range of rotation frequencies. 

\subsection{RMS radius of dipolar BEC in OL}

To understand the variation in the number of vortices with OL depth, we calculate the rms radius ($\langle r \rangle$) in the absence of rotation with different contact interaction strengths for a range of OL depths, $V_0 \in (0,60)$. 
\begin{figure*}[!ht]
\begin{center}
\includegraphics[width=0.80\linewidth,clip]{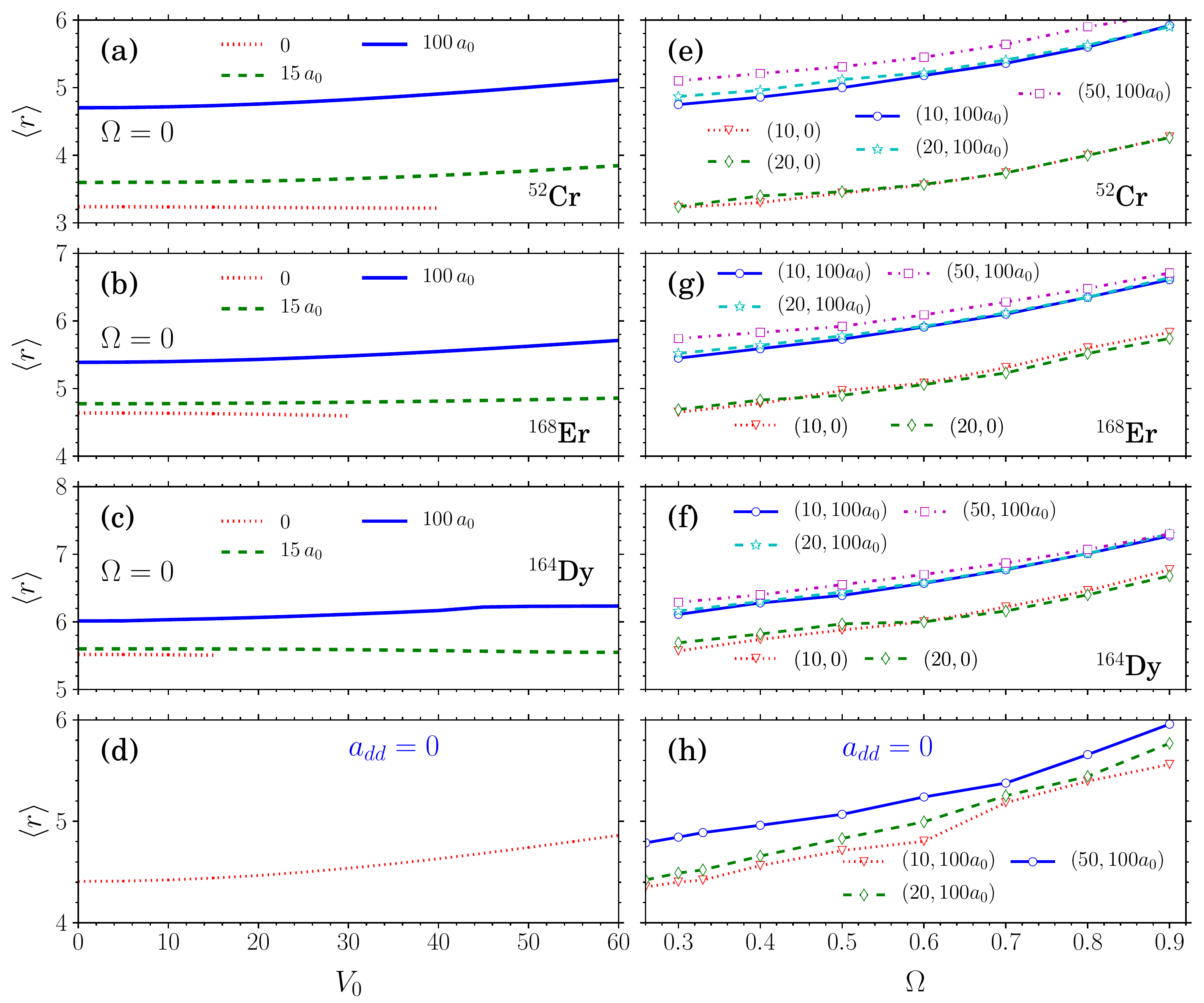}
\end{center}
\caption{(Color online) Numerically calculated rms radii of (a) $^{52}$Cr, (b) $^{168}$Er and (c) $^{164}$Dy {(d) non-dipolar BECs} for a range of OL depths, $V_0 \in (0, 60)$, with $\Omega=0$ for different contact interaction strengths, $a = 0$, $15 a_0$ and $100 a_0$. Figures~(e)-(h) depict the variation of rms radii of $^{52}$Cr, $^{168}$Er, $^{164}$Dy and {non-dipolar BECs}, respectively, against rotation frequency, $\Omega$, for different OL depths and contact interaction strengths, $(V_0, a)$.}
\label{fig6}
\end{figure*}
In Figures~\ref{fig6}(a)-\ref{fig6}(c), we plot $\langle r \rangle$ as a function of  $V_0$ for BECs of $^{52}$Cr, $^{168}$Er and $^{164}$Dy atoms in the absence of rotation ($\Omega = 0$) with contact interaction strengths $a = 0$, $15 a_0$ and $100 a_0$. When $a = 0$, the radii remain almost constant up to a critical depth of the OL and then the condensate collapses (dotted/red lines in Figures~\ref{fig6}(a)-\ref{fig6}(c)). Due to the constancy of the rms radius it could not accommodate more number of vortices. For $a = 15 a_0$, the rms radius (dashed/green lines) increases slightly with an increase of OL depth for $^{52}$Cr and $^{168}$Er BECs, and remains almost constant for $^{164}$Dy BEC. The condensate remains stable in all these cases. Similarly, for $a=100a_0$, the rms radius increases for $^{52}$Cr and $^{168}$Er BECs as shown in Figures~\ref{fig6}(a) and \ref{fig6}(b), respectively, by solid/blue lines. However, the rms radius for $^{164}$Dy BEC  increases up to $V_0 \approx 50$ and remains constant there after as shown in Figure~\ref{fig6}(c) by solid/blue line. 

Next we study the dependence of rms radius on $\Omega$ for different OL depths and contact interaction strengths. In Figures~\ref{fig6}(d)-\ref{fig6}(f), we plot the rms radii as a function of $\Omega$ for $^{52}$Cr, $^{168}$Er and $^{164}$Dy BECs, respectively, for different $(V_0,a)$, namely, $(10, 0)$, $(20, 0)$, $(10, 100 a_0)$, $(20, 100 a_0)$, and $(50, 100 a_0)$. When $a=100 a_0$, the rms radius increases as $\Omega$ and OL depth whereas it shrinks with the increase of dipolar strength. On the other hand, when $a=0$, there is a slight increase in the radii with respect to OL depth for $\Omega \lesssim 0.5$. However, for $\Omega > 0.5$ the radius shrinks as the depth of the OL increases. These are visible in the case of $^{168}$Er and  $^{164}$Dy BECs and is shown in Figures~\ref{fig6}(e) and \ref{fig6}(f) by dotted/red line with down triangle symbols and dashed/green line with diamond symbols. Thus, for stronger dipolar BEC in OL, the rms radius decreases when it rotates faster and hence the number of vortices also decreases at higher rotation frequencies. We have calculated the rms radius of non-dipolar BECs where it increases with respect to OL depth as shown in Figure~\ref{fig6}(d) and \ref{fig6}(h).  

\section{Summary and Conclusion}

\label{sec:conclusion}

We have studied the formation of vortices in rotating dipolar Bose-Einstein condensates in optical lattice by numerically solving the time-dependent Gross-Pitaevskii equation in two dimensions. Particularly, we have explored the influence of dipole-dipole interaction on the vortex nucleation, critical rotation frequency, time taken for the creation of vortices, number of vortices, and vortex structures in rotating dipolar BECs of $^{52}$Cr, $^{168}$Er and $^{164}$Dy atoms in OL. The critical rotation frequency has been found to decrease with an increase in the OL depth. The time taken for the nucleation of vortices is also estimated  and it has been observed that shallow OL can nucleate the vortices very rapidly in strongly dipolar BEC. The supplementary symmetry breaking due to OL in complement with the dipolar interaction accelerates the rapid creation of steady state vortices. Further, we have noticed phase transitions in the vortex structures due to the dipolar strengths and witnessed a regular pentagon structure with one vortex at center in $^{52}$Cr, square and triangular vortex lattice structures with slight distortion in $^{168}$Er and $^{164}$Dy condensates, respectively. We have also calculated the number of vortices as a function of OL depth, dipolar and contact interaction strengths, and rotation frequency. It has been shown that the number of vortices enhances at lower and moderate rotation frequencies while it gets reduced at higher rotation frequencies, which is evident from the calculation of the rms radius with respect to rotation frequency. It has been noted that dipolar BEC in OL with strong dipole-dipole interactions shrinks at higher rotation frequencies. 

\section*{Acknowledgments}

RKK acknowledges the support from Third World Acad\-emy of Sciences (TWAS) and Conselho Nacional de Desenvolvimento Cient\'{\i}fico e Tecnol\'ogico (CNPq), Brazil for the financial support in the form of TWAS-CNPq fellowship. The work of PM forms a part of  Council of Scientific and Industrial Research (CSIR Ref. No. 03(1186)/10/EMR-II), Department of Science and Technology (DST Ref. No. INT/FRG/DAAD/P-220/2012) both Government of India funded research projects and UGC-Special Assistance Programme.

\end{document}